\documentclass[12pt]{article}

\usepackage{a4}
\usepackage{floatfig,epsfig}
\usepackage{rotating}

\newcounter{my}

\newcommand{\la}[1]{\label{#1}}
\newcommand{\re}[1]{\ (\ref{#1})}
\newcommand{\nn}{\nonumber}
\newcommand{\ed}{\end{document}}
\newcommand{\be}{\begin{equation}}
\newcommand{\ee}{\end{equation}}

\newcommand{\ba}{\begin{eqnarray}}
\newcommand{\ea}{\end{eqnarray}}
\newcommand{\baz}{\begin{eqnarray*}}
\newcommand{\eaz}{\end{eqnarray*}}
\newcommand{\bb}{\begin{thebibliography}}
\newcommand{\eb}{\end{thebibliography}}
\newcommand{\ct}[1]{${\cite{#1}}$}

\newcommand{\bi}[1]{\bibitem{#1}}

\begin{document}

\initfloatingfigs
\sloppy
\thispagestyle{empty}

\vspace{1cm}

\mbox{}

\vspace{2cm}

\begin{center}
{\Large \bf  Instantons and Single Spin
Asymmetries}
\footnote{ The talk presented at the
Workshop "Physics with Polarized Protons at HERA",
August 1997, DESY-Zeuthen}\\[2cm]
{\large N.I.Kochelev}\\[0.2cm]
{\it
 Bogoliubov Laboratory of Theoretical Physics,\\
Joint Institute for Nuclear Research,\\
RU-141980 Dubna, Moscow region, Russia}\\ [0.3cm]

\end{center}
\vspace{2cm}

\begin{abstract}
\noindent
A new mechanism for single spin asymmetries in strong interaction
is proposed. It is based on instanton--induced nonperturbative 
chromo\-magnetic quark--gluon
interaction. We estimate the contribution of this interaction to
single spin asymmetry in quark--quark scattering.
It is shown, that interference between quark spin--flip amplitude
induced by instantons and spin--nonflip amplitude, which is related to
nonperturbative two gluon exchange, gives a large value for
a single spin asymmetry. 
Due to strong increasing of the instanton--induced spin--flip
amplitude with energy, this asymmetry does not vanish at high energies.
\end{abstract}
\newpage
\section{Introduction}
\vspace{1mm}
\noindent
In the recent years the interest in polarized hadron--hadron and
lepton--hadron
interactions at high energies has grown enormously.
 This interest stems from sensational result of the measurement by the
 EMC(CERN) Collaboration \ct{EMC} in polarized DIS of a part of 
 proton spin carried by  quarks.
 It has been shown that its value is very small.
 As a result, the "spin crisis" of a naive perturbative
QCD picture   of  spin--dependent lepton-hadron cross sections
 (see review \ct{rev}) arose.

However, it should be mentioned that a number of anomalous polarized
phenomena in hadron--hadron interaction have been found before
the EMC result \ct{Kri}. One of the very interesting results
was connected with very large single spin asymmetries
at  high energy  which was
observed by the E--704 Collaboration \ct{E704} in the
inclusive $\pi$ mesons production in the reactions $pp\rightarrow \pi X$,
$\bar pp\rightarrow \pi X$ by using transverse polarized proton and
antiproton beams. To clarify the mechanism of the single spin
asymmetries, the same experiment at more large energies has been included
 into  the programs of the RHIC--Spin Collaboration at Brookhaven
\ct{RHIC}
and HERA--$\vec N $ project  at DESY \ct{HN}.

It is very important to have the predictions 
of the single spin asymmetries based on  QCD.
As it is well--known, the quark-quark scattering
amplitude should contain a large spin--flip part  
to produce a substantial  quark single spin asymmetry.
The perturbative QCD predicts a decreasing of
spin-flip  effects with energy, therefore
  pQCD  fails to describe  the  data \ct{Kri} that show approximately
energy independence of single spin asymmetries.

Recently, a new approach to the theory of polarized effects based upon
nonperturbative QCD was proposed \ct{DorKoch} (see also some
applications of the instanton model to polarized DIS in\ct{Forte}).
In this approach the origin of the large polarized effects is the
interaction  of quarks with strong vacuum fluctuations of the gluon fields,
so-called instantons~~\ct{Pol}.

In the recent paper \ct{KOCH} it was shown that 
instantons lead to a new type of the nonperturbative
interaction, so--called chromomagnetic  quark--gluon interaction.
In \ct{koch2}  it was demonstrated that the contribution of
this interaction to  spin--dependent
$g_1(x,Q^2)$ structure function is large and allow us to
explain a most part of observed violation of the
Ellis--Jaffe sum rule.

 The fundamental feature of nonperturbative chromomagnetic quark--gluon
 interaction  is its  anomalous dependence on quark helicities.
So, this interaction leads to the quark helicity flip, just
as a usual perturbative quark--gluon  vertex conserves
the quark helicity.
Therefore this nonperturbative interaction
can be used as fundamental QCD mechanism
to explain very large single spin asymmetries observed in hadron--hadron
interaction.

The main goal of this article is to calculate the contribution
of  spin--flip chromomagnetic quark--gluon interaction induced by
instantons to quark single spin asymmetries.

\section{Quark--Gluon Chromomagnetic Interaction and Quark Single
Spin Asymmetry}
\vspace{1mm}
\noindent
We will consider the case when the  momentum transfer between quarks
is small
$p_\bot^2\ll S$, where $S=(p_1+p_2)^2$, $t=(p_1^\prime-p_1)^2=-p_\bot^2$ 
\footnote{ The generalization of this approach to the case of 
high transfer
momentum in quark--quark scattering  will be discussed elsewhere.}.
In this case the main contribution
to quark--quark scattering amplitude gives
Landshoff--Nachtmann (LN) pomeron which is based on the
exchange of a pair of nonperturbative gluons \ct{LN} (Fig.1a).
\begin{figure}[htb]
\centering
\epsfig{file=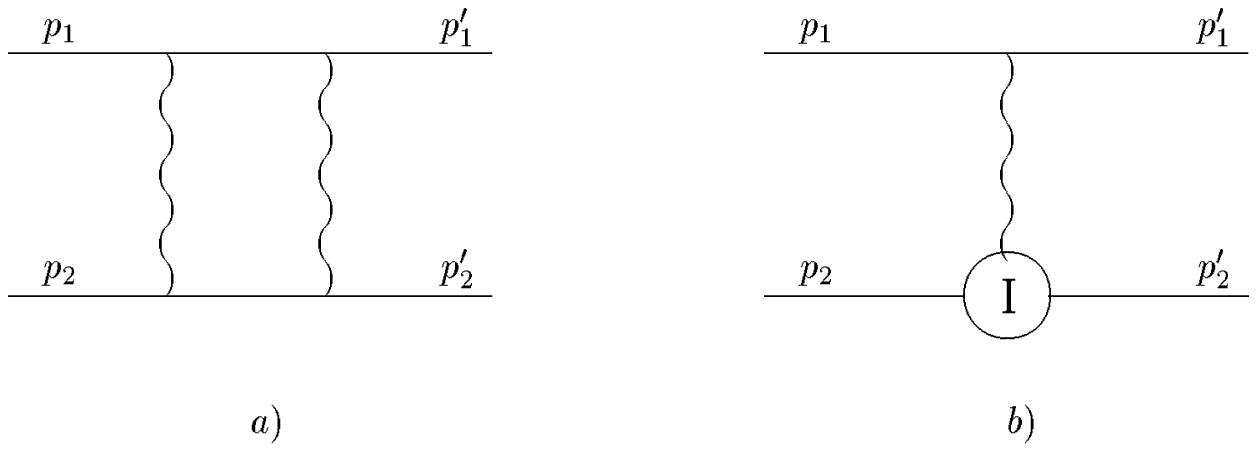,width=8cm}
\vskip 3cm
\caption{\it The  contribution to the  quark--quark scattering
amplitude: $a)$ the contribution from nonperturbative double
gluon exchange; $b)$ the contribution from chromomagnetic
quark--gluon interaction induced by instantons.
 The label $I$ denotes
instanton.}
\end{figure}

In the framework of this model the nonperturbative gluon
propagator reads
\ba
g^2D(p_\bot^2)=-\beta_0ae^{-p_\bot^2/\mu_0^2},
\la{prop}
\ea
where  $a=(4\pi^3)^{0.1}/\mu_0$,
$\beta_0\approx 2 GeV^{-1}$, and $\mu_0\approx 1.1 GeV $.

The double gluon  exchange with the perturbative quark-gluon
interaction does not lead to quark spin-flip and gives
rise to
helicity amplitudes  \ct{am} $\Phi_1=\Phi_3\approx M_{+,+;+,+}$.
In the framework of LN model these amplitudes are \ct{LN}
\be
\Phi_1=\Phi_3=i2\beta^2(p_\bot^2)S,
\la{nonflip}
\ee
where $\beta^2(p_\bot^2)=\beta_0^2exp(-p_\bot^2/2\mu_0^2)$.

The additional contribution to the quark--quark scattering amplitude is
related to the nonperturbative chromomagnetic  quark--gluon
interaction induced
by instantons \ct{KOCH} (see Fig.1b)
\be
\Delta {\cal L_A}=
-i\mu_a
\sum_q\frac{g}{2m_q^*}\bar q\sigma_{\mu\nu}
t^a qG_{\mu\nu}^a,
\label{e4}
\ee
where
$m_q^*=2\pi^2\rho_c^2<0\mid \bar qq\mid 0>/3$ is a quark mass in
the instanton vacuum.

The value of the quark anomalous chromomagnetic moment
in the liquid instanton model \ct{a5} is
\be
\mu_a=-\frac{f\pi}{2\alpha_s},
\label{a6}
\ee
 where $f=n_c\pi^2\rho_c^4$ is the so--called packing fraction of instantons
 in  vacuum.
The value of $n_c$ is connected with the value of the
gluon condensate by the formula
\be
n_c=<0\mid \alpha_sG_{\mu\nu}^a G_{\mu\nu}^a\mid 0>/16\pi
\approx 7.5~10^{-4}{\  }GeV^4.
\nn
\ee
The following estimate for
 the value of the anomalous quark chromomagnetic
moment has been obtained for $\rho_c=1.6GeV^{-1}$ in \ct{KOCH}
\begin{equation}
\mu_a=-0.2.
\nn
\ee
The quark-gluon chromomagnetic interaction gives rise to
 $\Phi_5=M_{++;+-}$ spin--flip amplitude
and straightforward calculation leads to the result
\be
\Phi_5=\frac{4\mu_ap_\bot g^2D(p_\bot^2)S}{9m_q^*}F(p_\bot^2\rho_c^2),
\la{flip}
\ee
where F(z) is the instanton form factor (see for example \ct{forte})
\be
F(z)=\frac{4}{z}(1-\frac{z}{2}K_2(\sqrt{z})).
\la{form}
\ee
\begin{figure}[htb]
\centering
\epsfig{file=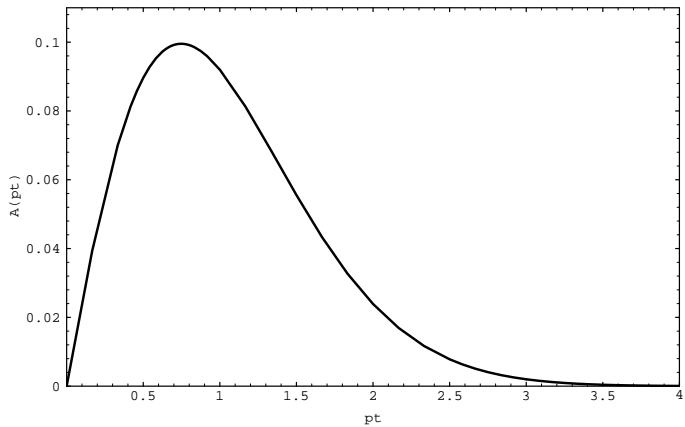,width=9cm}
\vskip -3cm
\caption{\it The instanton contribution to  the quark single
spin asymmetry.}
\end{figure}
The single spin asymmetry on the quark level can be written in the
following  form
\be
A=\frac{2Im(\Phi_5(\Phi_1+\Phi_3))}{|\Phi_1|^2+|\Phi_3|^2+4|\Phi_5|^2},
\la{as}
\ee
where we have neglected the contribution which comes from double spin-flip
amplitudes $\Phi_2$ and $\Phi_4$, that are connected with
high order instanton contributions.

The result of the calculation of the quark single spin asymmetry
is presented in Fig.2
The value of the asymmetry is rather large $A\approx 10\%$
at $p_\bot\approx 1GeV$. The magnitude of the asymmetry
strongly depends on the value of the average size
of instantons in QCD vacuum. For example,  for
slightly different  value  $\rho_c=2 GeV^{-1}$ \ct{a5}, the final
result for the asymmetry is approximately $1.5$ of the original value.
The above calculation, which was performed by using nonperturbative
LN gluon propagator \re{prop}, is valid only at rather small value
of transfer momentum $p_\bot\leq 1 GeV$. At larger values of
$p_\bot$ one should use a perturbative gluon propagator instead of
nonperturbative one \re{prop} and take into
account the instanton contribution to spin-nonflip amplitude as well.

It should be mentioned also that due to the
similar growth of both instanton--induced spin--flip amplitude \re{flip}
 and of spin-nonflip LN amplitude \re{nonflip} with energy, 
the single spin asymmetry turns out to be energy--independent. This result
is supported
by the experimental data on single spin asymmetries \ct{Kri} and
is opposite to the prediction
 of  perturbative QCD, $A\approx m_q/\sqrt{S}$ \ct{pqcd}.

\section{Summary}
\vspace{1mm}
\noindent
In summary,  the instanton--induced chromomagnetic quark--gluon
interaction
leads  to a large quark single  spin asymmetry at high energy.
The magnitude  of the single spin asymmetry is determined by
the parameters of the instantons in the QCD vacuum. Therefore
the investigation of the single spin asymmetries can give very
important information on the structure of the QCD vacuum.

\section*{Acknowledgements}
\vspace{1mm}
\noindent
The author is
thankful to M.Anselmino, A.E.Dorokhov,
A.V.Efremov, Ph.Ratcliffe, E.Leader, and W.--D.Nowak  for helpful
discussions.

This work was supported in part by the Heisenberg--Landau program and by the
Russian Foundation for Fundamental Research (RFFR) 96--02--18096.

\end{document}